\documentclass[12pt,letterpaper]{article}
\pdfoutput=1
\usepackage{jheppub}
\usepackage{amsfonts, amsthm}
\usepackage[english]{babel}
\usepackage[utf8]{inputenc}
\usepackage{slashed}
\usepackage{mathrsfs}
\hypersetup{unicode}

\newcommand{\eq}{\begin{equation}}
\newcommand{\feq}{\end{equation}}
\newcommand{\eqn}{\begin{eqnarray}}
\newcommand{\feqn}{\end{eqnarray}}

\newcommand{\ma}[1]{\mbox{$\mathcal{#1}$}}

\newcommand{\masf}[1]{\mbox{$\mathsf{#1}$}}

\newcommand{\I}{\ma{I}}
\newcommand{\R}{\ma{R}}
\newcommand{\kk}{\masf{k}}
\newcommand{\KK}{\masf{K}}
\newcommand{\A}{\masf{A}}
\newcommand{\PP}{\masf{P}}

\title{Supersymmetric black holes and attractors in gauged supergravity with hypermultiplets}
 
\author{Samuele Chimento, Dietmar Klemm and Nicol\`o Petri}
\affiliation{Dipartimento di Fisica, Universit\`a di Milano, and \\
INFN, Sezione di Milano, \\
Via Celoria 16, 20133 Milano, Italy.
}
\emailAdd{samuele.chimento@mi.infn.it}
\emailAdd{dietmar.klemm@mi.infn.it}
\emailAdd{nicolo.petri@mi.infn.it}
\preprint{IFUM-1038-FT}

\abstract{We consider four-dimensional $N=2$ supergravity coupled to vector- and hypermultiplets,
where abelian isometries of the quaternionic K\"ahler hypermultiplet scalar manifold are gauged.
Using the recipe given by Meessen and Ort\'{\i}n in arXiv:1204.0493, we analytically construct
a supersymmetric black hole solution for the case of just one vector multiplet with prepotential
${\cal F}=-i\chi^0\chi^1$, and the universal hypermultiplet. This solution has a running dilaton,
and it interpolates between 
$\text{AdS}_2\times\text{H}^2$ at the horizon and a hyperscaling-violating type geometry at infinity,
conformal to $\text{AdS}_2\times\text{H}^2$. It carries two magnetic charges that are completely
fixed in terms of the parameters that appear in the Killing vector used for the gauging.

In the second
part of the paper, we extend the work of Bellucci et al.~on black hole attractors in gauged supergravity to
the case where also hypermultiplets are present. The attractors are shown to be governed by an
effective potential $V_{\text{eff}}$, which is extremized on the horizon by all the scalar fields of the
theory. Moreover, the entropy is given by the critical value of $V_{\text{eff}}$. In the limit of
vanishing scalar potential, $V_{\text{eff}}$ reduces (up to a prefactor) to the usual black hole potential.
}


\begin{document}
\maketitle
\flushbottom

\section{Introduction}

Black holes in gauged supergravity theories provide an important testground
to address fundamental questions of gravity, both at the classical and quantum level. Among these are
for instance the problems of black hole microstates, the final state of black hole evolution,
uniqueness- or no hair theorems, to mention only a few of them. In gauged supergravity, the solutions
typically have AdS asymptotics, and one can then try to study these issues guided by the AdS/CFT
correspondence.
On the other hand, black hole solutions to these theories are also relevant for a number of recent 
developments in high energy- and especially in condensed matter physics, since they provide the
dual description of certain condensed matter systems at finite temperature, cf.~\cite{Hartnoll:2009sz}
for a review. In particular, models that contain Einstein gravity coupled to $\text{U}(1)$ gauge fields\footnote{The necessity of a bulk $\text{U}(1)$ gauge field arises, because a basic ingredient of
realistic condensed matter systems is the presence of a finite density of charge carriers.}
and neutral scalars have been instrumental to study transitions from Fermi-liquid to
non-Fermi-liquid behaviour, cf.~\cite{Charmousis:2010zz,Iizuka:2011hg} and references therein.
In AdS/condensed matter applications one is often interested in including a charged scalar
operator in the dynamics, e.g.~in the holographic modeling of strongly coupled
superconductors \cite{Hartnoll:2008vx}. This is dual to a charged scalar field in the bulk, that
typically appears in supergravity coupled to gauged hypermultiplets. It would thus be desirable to
dispose of analytical black hole solutions to such theories. In the first part of the
present paper we will make a first step
in this direction. Solving the corresponding second order equations of motion is generically quite
involved, such that one is forced to resort to numerical techniques. For this reason we shall look here
for BPS black holes, which satisfy first order equations, and make essential use of the
results of \cite{Meessen:2012sr}, where all supersymmetric backgrounds of $N=2$, $d=4$ gauged
supergravity coupled to both vector- and hypermultiplets were classified. This provides
a systematic method to obtain BPS solutions, without the necessity to guess some suitable ans\"atze.
Let us mention here that black holes in four-dimensional gauged supergravity with hypers were also
obtained numerically in \cite{Halmagyi:2013sla}. Solutions that have ghost modes (i.e., with at
least one negative eigenvalue of the special K\"ahler metric) were constructed in \cite{Hristov:2010eu}.
In five dimensions, a singular solution of supergravity with gauging of the axionic shift symmetry of the universal hypermultiplet was derived in \cite{Gutperle:2001vw}. Finally, ref.~\cite{Erbin:2014hsa} analyzed
the near-horizon geometries of static BPS black holes in four-dimensional $N=2$ supergravity with
gauging of abelian isometries of the hypermultiplet scalar manifold, while the authors of
\cite{Halmagyi:2011xh} found nonrelativistic (Lifshitz and Schr\"odinger) solutions in the same theory
for the canonical example of a single vector- and a single hypermultiplet\footnote{For related work
cf.~\cite{Cassani:2011sv}, where Lifshitz solutions in general $N=2$, $d=4$ supergravity models
were obtained by reducing $d=5$ theories with AdS vacua.}.

Another point of interest addressed in this paper is the attractor mechanism \cite{Ferrara:1995ih,
Strominger:1996kf,Ferrara:1996dd,Ferrara:1996um,Ferrara:1997tw}, that has been the subject
of extensive research in the asymptotically flat case, but for which not very much has been done for
black holes with more general asymptotics. First steps towards a systematic analysis of
the attractor flow in gauged supergravity were made in \cite{Morales:2006gm,Bellucci:2008cb} for the
non-BPS and in \cite{Huebscher:2007hj,Cacciatori:2009iz,Dall'Agata:2010gj,Kachru:2011ps} for the
BPS case. Some interesting results have been found, for instance the appearance of flat directions
in the effective black hole potential for BPS flows \cite{Cacciatori:2009iz}, a property that does not
occur in ungauged $N=2$, $d=4$ supergravity \cite{Ferrara:1997tw}, at least as long as the metric
of the scalar manifold is strictly positive definite.

In the second part of our paper we extend the work of \cite{Bellucci:2008cb} to include also gauged
hypermultiplets. We shall construct an effective potential $V_{\text{eff}}$ that depends on both the
usual black hole potential and the potential for the scalar fields. $V_{\text{eff}}$ governs the attractors,
in the sense that it is extremized on the horizon by all the scalar fields of the theory, and the entropy is
given by the critical value of $V_{\text{eff}}$. As in \cite{Bellucci:2008cb}, our analysis does not make use
of supersymmetry, so our results are valid for any static extremal black hole in four-dimensional $N=2$
matter-coupled supergravity with gauging of abelian isometries of the hypermultiplet scalar manifold.

The remainder of this paper is organized as follows: In the next section, we
briefly review $N=2$, $d=4$ gauged supergravity coupled to vector- and
hypermultiplets. Section \ref{sec:susy-sol} summarizes the general recipe
to construct supersymmetric solutions provided in \cite{Meessen:2012sr}. In \ref{sec:bhsol},
a simple model is considered that has just one vector multiplet with special K\"ahler prepotential
${\cal F}=-i\chi^0\chi^1$, and the universal hypermultiplet. In this setting, the equations of
\cite{Meessen:2012sr} are then solved and a genuine BPS black hole with running dilaton and two
magnetic charges is constructed. Section \ref{sec:attractor} contains an extension of the results
of \cite{Bellucci:2008cb} on black hole attractors in gauged supergravity to the case that includes also
hypermultiplets. Section \ref{sec:final} contains our conclusions and some final remarks.

\section{Matter-coupled \texorpdfstring{$N=2$, $d=4$}{N=2, d=4} gauged supergravity\label{sugra}}

The gravity multiplet of $N=2$, $d=4$ supergravity can be coupled to a number $n_V$ of vector multiplets and to $n_H$ hypermultiplets. The 
bosonic sector then includes the vierbein $e^a{}_\mu$, $\bar n\equiv n_V+1$ vector fields $A^\Lambda{}_\mu$ with $\Lambda=0,\dots n_V$ (the
graviphoton plus $n_V$ other fields from the vector multiplets), $n_V$ complex scalar fields $Z^i$, $i=1,\dots,n_V$, and $4 n_H$ real 
hyperscalars $q^u$, $u=1,\dots,4 n_H$.

The complex scalars $Z^i$ of the vector multiplets parametrize an $n_V$-dimensional special K\"ahler manifold, i.e. a K\"ahler-Hodge 
manifold, with K\"ahler metric $\ma{G}_{i\bar\jmath}(Z,\bar Z)$, which is the base of a symplectic bundle with the covariantly holomorphic 
sections\footnote{The conventions and notation used in this paper are those of refs. \cite{Meessen:2012sr,Bellorin:2005zc}}.
\begin{equation}
 \ma{V}=\left(\begin{array}{c}
                   \ma{L}^\Lambda\\
                   \ma{M}_\Lambda
                  \end{array}\right), \qquad 
                  \ma{D}_{\bar \imath}\ma{V}\equiv\partial_{\bar \imath}\ma{V}
                  -\frac{1}{2}\left(\partial_{\bar \imath}\ma{K}\right)\ma{V}=0\,,
\end{equation}
obeying the constraint
\begin{equation}
 \left\langle\ma{V}|\ma{\bar V}\right\rangle\equiv
 \ma{\bar L}^\Lambda\ma{M}_\Lambda-\ma{L}^\Lambda \ma{\bar M}_\Lambda=-i\,, \label{eq:sympcond}
\end{equation}
where \ma{K} is the K\"ahler potential.
Alternatively one can introduce the explicitly holomorphic sections of a different symplectic bundle,
\begin{equation}
 \Omega\equiv e^{-\mathcal{K}/2}\ma{V}\equiv\left(\begin{array}{c}
						  \chi^\Lambda\\
						  \ma{F}_\Lambda
						  \end{array}\right)\,.
\end{equation}
In appropriate symplectic frames it is possible to choose a homogeneous function of second degree $\ma{F}(\chi)$, called prepotential, such
that $\ma{F}_\Lambda=\partial_\Lambda \ma{F}$.
In terms of the sections $\Omega$ the constraint (\ref{eq:sympcond}) becomes
\begin{equation}
\label{eq:sympcond2}
 \left\langle\Omega|\bar{\Omega}\right\rangle\equiv\bar{\chi}^\Lambda\ma{F}_\Lambda-\chi^\Lambda{\ma{\bar F}}_\Lambda=
 -i e^{-\mathcal{K}}.
\end{equation}
The couplings of the vector fields to the scalars are determined by the $\bar n\times\bar n$ period matrix \ma{N}, defined by the 
relations
\begin{equation}
 \ma{M}_\Lambda = \ma{N}_{\Lambda\Sigma}\,\ma{L}^\Sigma, 
 \qquad \ma{D}_{\bar\imath}\ma{\bar M}_\Lambda=\ma{N}_{\Lambda\Sigma}\,\ma{D}_{\bar \imath}\ma{\bar L}^\Sigma\,.
\end{equation}
If the theory is defined in a frame in which a prepotential exists, \ma{N} can be obtained from
\begin{equation}
  \label{eq:period_matrix_prep}
  \ma{N}_{\Lambda\Sigma}=\ma{\bar F}_{\Lambda\Sigma}
  +2 i\frac{\left(N_{\Lambda\Gamma}\chi^\Gamma\right)\left(N_{\Sigma\Delta}\chi^\Delta\right)}{\chi^\Omega N_{\Omega\Psi}\chi^\Psi}\,,
\end{equation}
where $\ma{F}_{\Lambda\Sigma}=\partial_\Lambda\partial_\Sigma \ma{F}$ and $N_{\Lambda\Sigma}\equiv\mathfrak{Im}(\ma{F}_{\Lambda\Sigma})$.

The $4 n_H$ real hyperscalars $q^u$ parametrize a quaternionic K\"ahler manifold with metric $\masf{H}_{uv}(q)$.
A quaternionic K\"ahler manifold is a $4n$-dimensional Riemannian manifold admitting a locally defined triplet 
${\vec \KK}_u^{\phantom{u}v}$ of almost complex structures satisfying the quaternion relation
 \begin{equation}
  \KK^1 \KK^2=\KK^3\,,
 \end{equation}
and whose Levi-Civita connection preserves $\vec \KK$ up to a rotation,
 \begin{equation}
  \label{eq:quaternionic_kahler_complexstruct_rotation}
  \nabla_w {\vec \KK}_u^{\phantom{u}v}+\,\vec\A_w\times{\vec \KK}_u^{\phantom{u}v}=0\,,
 \end{equation}
with SU(2) connection $\vec\A\equiv \vec\A_u (q)\, dq^u$. An important property is that the SU(2) curvature is proportional to the
complex structures,
 \begin{equation}
  \label{eq:quaternionkahl_su2curv}
  \masf{F}^x\equiv\, d\A^x+\frac12\varepsilon^{xyz}\A^y\wedge\A^z=-2\,\KK^x\,.
 \end{equation}
 
We will only consider gaugings of abelian symmetries of the action. Under the action of abelian symmetries, the complex scalars $Z^i$ 
transform trivially, so that we will be effectively gauging abelian isometries of the quaternionic-K\"ahler metric $\masf{H}_{uv}$.
These are generated by commuting Killing vectors $\kk_\Lambda{}^u (q)$, $[k_\Lambda,k_\Sigma]=0$, and the requirement that the 
quaternionic-K\"ahler structure is preserved implies the existence of a triplet of Killing prepotentials, or moment maps, 
$\PP_\Lambda{}^x$ for each Killing vector such that
\begin{equation}
\label{eq:mommaps}
 \PP_\Lambda{}^x=\frac{1}{2n_H}\KK^x{}_u{}^v\nabla_v\kk_\Lambda{}^u\,,\qquad 
 \masf{D}_u\PP_\Lambda{}^x\equiv \partial_u\PP_\Lambda{}^x+\varepsilon^{xyz}\A_{\phantom{y} u}^y\PP_\Lambda{}^z=
                                                                                   -2\,\KK^x{}_{uv}\kk_\Lambda{}^v\,.
\end{equation}

The bosonic action reads
\begin{multline}
\label{eq:action}
  S = \int d^{4}x \sqrt{|g|}\left[R+2\,\ma{G}_{i\bar\jmath}\,\partial_{\mu}Z^{i}\partial^{\mu}\bar Z^{\bar\jmath}
    +2\, \masf{H}_{uv}\, \mathfrak{D}_{\mu} q^{u} \mathfrak{D}^{\mu} q^{v} \right.\\
    \left.+2\,I_{\Lambda\Sigma}\, F^{\Lambda\, \mu\nu}F^{\Sigma}{}_{\mu\nu} 
    -2\, R_{\Lambda\Sigma}\,F^{\Lambda\, \mu\nu}\star F^{\Sigma}{}_{\mu\nu}-V(Z,\bar Z,q)\right]\,,
\end{multline}
where the scalar potential has the form
\begin{equation}
 \label{eq:scal_pot}
 V(Z,\bar Z,q) = 
  g^2\left[2\ma{\bar L}^\Lambda\ma{L}^\Sigma(\masf{H}_{uv}\kk_\Lambda{}^u \kk_\Sigma{}^v-\PP_\Lambda{}^x\PP_\Sigma{}^x)
  -\frac14 I^{\Lambda\Sigma}\PP_\Lambda{}^x \PP_\Sigma{}^x\right]\,,
\end{equation}
the covariant derivatives acting on the hyperscalars are
\begin{equation}
 \mathfrak{D}_\mu q^u=\partial_\mu q^u+gA^\Lambda{}_\mu\kk_\Lambda{}^u\,,
\end{equation}
and
\begin{equation}
I_{\Lambda\Sigma}\equiv\mathfrak{Im}(\ma{N}_{\Lambda\Sigma})\,,\qquad R_{\Lambda\Sigma}\equiv\mathfrak{Re}(\ma{N}_{\Lambda\Sigma})\,,
\qquad I^{\Lambda\Sigma} I_{\Sigma\Gamma}=\delta^\Lambda{}_\Gamma\,.
\end{equation}


\section{Supersymmetric solutions}
\label{sec:susy-sol}

All the timelike supersymmetric solutions to $N=2$ gauged supergravity in four dimensions were characterized by Meessen and Ortín
in \cite{Meessen:2012sr}. Here we summarize their results, restricted to the case of abelian gauging.

The expressions and equations that follow are given in terms of bilinears constructed out of the Killing spinors,
\begin{equation}
X=\frac{1}{2} \varepsilon^{IJ}\bar\epsilon_I\epsilon_J\,, \qquad
V_a=i\bar\epsilon^I\gamma_a\epsilon_I\,, \qquad
V_a^x=i(\sigma^x)_I^{\phantom{I}J}\bar\epsilon^I\gamma_a\epsilon_J\,,\label{eq:bilinears}
\end{equation}
and of the real symplectic sections of K\"ahler weight zero
\begin{equation}
\R\equiv \mathfrak{Re}(\ma{V}/X)\,, \qquad
\I\equiv \mathfrak{Im}(\ma{V}/X)\,. \label{eq:risections}
\end{equation}
The metric and vector fields take the form
\begin{align}
 \label{eq:gensol_metric}
 ds^2&=2\left|X\right|^2(dt+\omega)^2-\frac{1}{2\left|X\right|^2}h_{mn}dy^mdy^n\,, \\
 &\nonumber\\
 \label{eq:gensol_gauge}
 A^\Lambda&=-\frac{1}{2}\mathcal{R}^\Lambda V + \tilde{A}^\Lambda_m dy^m\,,
\end{align}
where the 3-dimensional metric $h_{mn}$ must admit a dreibein $V^x$ satisfying the structure
equation\footnote{Eq.~\eqref{eq:3dcartan2} corrects a typo in \cite{Meessen:2012sr}; the terms
containing the moment maps must have the opposite sign w.r.t.~the one in \cite{Meessen:2012sr}.}
\begin{equation}
\label{eq:3dcartan2}
 dV^x+\epsilon^{xyz}\left(\A^y-g\tilde A^\Lambda\PP_\Lambda{}^y\right)\wedge V^z
 +\frac{g}{\sqrt{2}}\I^\Lambda\PP_\Lambda{}^y V^y\wedge V^x=0\,.
\end{equation}
$|X|^2$ can be determined from $\R$ and $\I$,
\begin{equation}
 \frac{1}{2\left|X\right|^2}=\left\langle\R|\I\right\rangle\,,
\end{equation}
the 1-form $V$ is given by
\begin{equation}
 V=2\sqrt{2}\left|X\right|^2(dt+\omega)\,,
\end{equation}
and the spatial 1-form $\omega$ satisfies
\begin{equation}
\label{eq:omega}
 (d\omega)_{xy}=2\,\varepsilon_{xyz}\left[\langle\I|\partial_{z}\I\rangle
-\frac{g}{2\sqrt{2}|X|^{2}}\R^\Lambda\PP_\Lambda{}^z\right]\,.
\end{equation}

The complex scalars $Z^i$, the sections $\R$ and $\I$, the 1-form $\omega$, the function $X$ and the hyperscalars $q^u$ are all 
time-independent. 

The complex scalars depend, in a way that depends on the chosen parametrization of the special K\"ahler manifold,
on the sections $\R$ and $\I$. A common simple choice of parametrization is $\chi^0=1$, $\chi^i=Z^i$, in which case
\begin{equation}
 Z^i=\frac{\ma{L}^i}{\ma{L}^0}=\frac{\R^i+i\,\I^i}{\R^0+i\,\I^0}\,.
\end{equation}
The effective 3-dimensional gauge connection $\tilde A^\Lambda$ must satisfy
\begin{equation}
\label{eq:ftilde}
(d\tilde A^\Lambda)_{xy}=\tilde{F}^\Lambda{}_{xy}=-\frac1{\sqrt{2}}\varepsilon_{xyz}(\partial_z
\I^\Lambda+g\ma{B}^\Lambda{}_z)\,,
\end{equation}
with
\begin{equation}
\ma{B}^\Lambda{}_x\equiv\sqrt{2}\left[\R^\Lambda\R^\Sigma+\frac{1}{8|X|^{2}} I^{\Lambda\Sigma}\right]\PP_\Sigma{}^x\,,
\end{equation}
from which follows the integrability condition
\begin{equation}
\label{eq:Iintegrab}
\tilde{\nabla}^{2}\I^\Lambda +g\tilde{\nabla}_x\ma{B}^\Lambda{}_x=0\,. 
\end{equation}
A similar condition holds for the $\I_\Lambda$'s,
\begin{equation}
\label{eq:iintegrab}
 \tilde{\nabla}^{2}\I_\Lambda +g\tilde{\nabla}_x\ma{B}_{\Lambda{}\,x}=
 \frac{g}{\sqrt{2}}\langle\I|\partial_x\I\rangle\ \PP_\Lambda{}^x 
+\ \frac{g^2}{4|X|^{2}}\R^\Sigma\left[\kk_{\Lambda\, u}\kk_\Sigma{}^{u}-\PP_\Lambda{}^x\PP_\Sigma{}^x\ \right]\,,
\end{equation}
where
\begin{equation}
\ma{B}_{\Lambda\, x}\equiv \sqrt{2}\left[\R_\Lambda\R^\Sigma +\frac{1}{8|X|^{2}}\ R_{\Lambda\Gamma}I^{\Gamma\Sigma}\right]\PP_\Sigma{}^x\,.
\end{equation}

Finally, the hyperscalars must satisfy the equation
\begin{equation}
\label{eq:hyperder}
 \KK^{x\, u}{}_{v}V^{x\, \mu}\mathfrak{D}_{\mu}q^{v}+\sqrt{2}g |X|^{2}\I^\Lambda\kk_\Lambda{}^{u}  = 0\,.
\end{equation}

For a given special geometric model the sections $\R$ can always, at least in principle, be determined in terms of the sections $\I$, by 
solving the so-called \emph{stabilization equations}. This means that to obtain a supersymmetric solution one needs to solve the above
equations for $\I^\Lambda$, $\I_\Lambda$, $\omega$, $V^x$ and $q^u$.

\section{A black hole solution}
\label{sec:bhsol}

We now turn to the task of obtaining an explicit solution with non-trivial hyperscalars. To do so, we consider a simple theory with
just one vector multiplet and one hypermultiplet, $n_V=n_H=1$.

More specifically, let the hypermultiplet be the \emph{universal hypermultiplet} \cite{Cecotti:1988qn}. The scalar fields in this 
multiplet, denoted by $(\phi, a, \xi^0, \xi_0)$, parametrize the quaternionic space
$\text{SU}(2,1)/\text{U}(2)$, with metric
\begin{equation}
\label{eq:hypersol_hyper_metric}
 \masf{H}_{uv}dq^u dq^v=d\phi^2+\frac14 e^{4\phi}\left(da-\frac12\langle\xi|d\xi\rangle\right)^2
 +\frac14 e^{2\phi}[(d\xi^0)^2+(d\xi_0)^2]\,,
\end{equation}
where $\langle\xi|d\xi\rangle\equiv\xi_0 d\xi^0-\xi^0 d\xi_0$,
and the corresponding SU(2) connection has components
\begin{equation}
 \A^1=e^\phi d\xi_0\,,\quad \A^2=e^\phi d\xi^0\,,\quad \A^3=\frac{e^{2\phi}}{2}\left( da-\frac12 \langle\xi|d\xi\rangle\right) \,.
\end{equation}
As for the vector multiplet, we choose a special geometric model specified by the prepotential
\begin{equation}
 \label{eq:hypersol_prepot}
 \ma{F}(\chi)=-i\chi^0\chi^1\,,
\end{equation}
with the parametrization $\chi^0=1$, $\chi^1=Z$. Then it is easy to obtain from \eqref{eq:sympcond2} the
K\"ahler potential $\ma{K}=-\log \left[ 4\,\mathfrak{Re}(Z) \right]$ and the K\"ahler metric
\begin{equation}
\label{eq:hypersol_kahl_metric}
 \mathcal{G}_{Z\bar Z} = \partial_Z\partial_{\bar Z}\ma{K}= \frac{1}{4 \,\mathfrak{Re}(Z)^2}\,,
\end{equation}
while the period matrix $\ma{N}_{\Lambda\Sigma}$, giving the scalar-vector couplings, is calculated from eq.~\eqref{eq:period_matrix_prep} 
to be
\begin{equation}
 \mathcal{N}=-i\left(\begin{array}{cc}
                    Z&0\\
                    0&\frac{1}{Z}
                   \end{array}\right)\,.
\end{equation}

Using the definition \eqref{eq:risections}, the dependence of the $\R$ section on the $\I$ section for this special geometric model is 
readily seen to be
\begin{equation}
\label{eq:hypersol_stab_sol}
 \R^0=-\I_1\,, \qquad \R^1=-\I_0\,, \qquad \R_0=\I^1\,, \qquad \R_1=\I^0\,,
\end{equation}
so that the complex scalar is given by
\begin{equation}
\label{eq:hypersol_Zscal}
 Z=\frac{\R^1+i\I^1}{\R^0+i\I^0}=\frac{\I_0-i\I^1}{\I_1-i\I^0}\,,
\end{equation}
and
\begin{equation}
 \frac{1}{2\left|X\right|^2}=\left\langle\R|\I\right\rangle=2\left( \I^0\I^1+\I_0\I_1 \right)\,.
\end{equation}
Since the theory includes two vector fields, we can choose to gauge up to two isometries of the metric $\masf{H}_{uv}$. We choose to gauge 
the (commuting) isometries generated by the Killing vectors
\begin{equation}
\label{eq:hypersol_killing_vecs}
 \kk_\Lambda=k_\Lambda\partial_a+\delta^0{}_\Lambda c \left( \xi_0\partial_{\xi^0}-\xi^0\partial_{\xi_0} \right)\,,
\end{equation}
where $k_\Lambda$ and $c$ are constants. This means that we are gauging the $\mathbb{R}$ group of the translations along $a$ with
the combination $A^\Lambda k_\Lambda$, and the U(1) group of rotations in the $\xi^0$--$\xi_0$ plane with the field $A^0$. \eqref{eq:hypersol_killing_vecs} is a subcase of the Killing vector considered
in \cite{Halmagyi:2013sla}, and corresponds to setting
\begin{equation}
Q_{\Lambda A} = {Q_{\Lambda}}^A = 0\,, \qquad \mathbb{U} = \left(\begin{array}{cc} 0 & c \\
-c & 0\end{array}\right)
\end{equation}
in eqs.~(3.8) and (3.9) of \cite{Halmagyi:2013sla}.
The triholomorphic moment maps associated with the Killing vectors \eqref{eq:hypersol_killing_vecs} can be obtained from 
\eqref{eq:mommaps}, and are
\begin{align}
 \PP_\Lambda{}^1&=-\delta^0{}_\Lambda c\, \xi^0 e^\phi\,,\qquad\qquad \PP_\Lambda{}^2=\delta^0{}_\Lambda c\, \xi_0 e^\phi\,,\nonumber\\
 &\label{eq:hypersol_tri_mommaps}\\
 \PP_\Lambda{}^3&=\delta^0{}_\Lambda c\left[1-\frac14 e^{2\phi}\left((\xi^0)^2+(\xi_0)^2\right)\right]+\frac12 k_\Lambda e^{2\phi}
 \nonumber\,.
\end{align}
With these choices the scalar potential \eqref{eq:scal_pot} reads
\begin{multline}
\label{eq:hypersol_scalpot}
 V=\frac{g^2}{2}\left\{\frac{1}{Z+\bar{Z}}\left[\frac{e^{4\phi}}{4}\left[k_0-\frac{c}{2}\left((\xi^0)^2+(\xi_0)^2 \right) \right]^2
    - c^2 - k_0 c\, e^{2\phi}  \right]\right.\\
    \left. +\frac{Z\bar Z}{Z+\bar Z}\frac{e^{4\phi}}{4} k_1^2 - k_1 c\,e^{2\phi}\right\}\,.
\end{multline}
For simplicity we will look for solutions with $\R^0=\R^1=\I_0=\I_1=0$, which implies from \eqref{eq:hypersol_Zscal} that the scalar $Z$ 
is real and from \eqref{eq:gensol_gauge} that the gauge fields are in a purely magnetic configuration. From eq.~\eqref{eq:omega} 
follows that $\omega$ is a closed 1-form, and can be reabsorbed by a redefinition of the coordinate $t$, leading to static solutions.
This choice also implies that eq.~\eqref{eq:iintegrab} is trivially satisfied.

We will also take the hyperscalar $a$ to be constant and $\xi^0=\xi_0=0$. Note that the scalar potential
\eqref{eq:hypersol_scalpot} has then a critical point at $Z=-k_0/k_1$ and $e^{2\phi}=-c/k_0$,
with $V_{\text{crit}}=3k_1g^2c^2/(8k_0)$. Since the absence of ghost modes requires $Z>0$, one
needs $k_0/k_1<0$ (and of course $c/k_0<0$) to have a critical point of the potential. With the choice
$\xi^0=\xi_0=0$, the moment maps \eqref{eq:hypersol_tri_mommaps} become
\begin{equation}
 \label{eq:hypersol_tri_mommaps2}
 \PP_\Lambda{}^1=\PP_\Lambda{}^2=0\,,\qquad \PP_\Lambda{}^3=\delta^0{}_\Lambda c+\frac12 k_\Lambda e^{2\phi}\,.
\end{equation}
Eq.~\eqref{eq:3dcartan2} implies then $dV^3=0$, hence there exists a function $r$ (that we will use as
a coordinate) such that locally 
\begin{equation}
\label{eq:hypersol_V3}
 V^3=dr\,.
\end{equation}
We shall impose radial symmetry on the solution by requiring the scalar fields $Z$, $\phi$ and the sections $\I^\Lambda$ to depend only on 
$r$. 

The $\phi$, $\xi^0$ and $\xi_0$ components of equation \eqref{eq:hyperder} reduce then to the constraint
\begin{equation}
\label{eq:hypersol_gauge_constr}
 A^\Lambda{}_x k_\Lambda=0 \qquad\Longrightarrow\qquad \tilde A^\Lambda k_\Lambda=0\,,
\end{equation}
while the $a$ component becomes
\begin{equation}
\label{eq:hypersol_derphi}
 \phi'=\frac{g}{2\sqrt{2}}\,e^{2\phi}\,\I^\Lambda k_\Lambda\,,
\end{equation}
where the prime stands for a derivative with respect to $r$.

If we now introduce the remaining coordinates $\vartheta$ and $\phi$ by choosing
\begin{equation}
 \label{eq:hypersol_V1_V2}
 V^1=e^{W(r)}d\vartheta\,, \qquad V^2=e^{W(r)}f(\vartheta) d\varphi\,,
\end{equation}
where at this stage $f$ is an arbitrary function of $\vartheta$, the remaining components of eq.~\eqref{eq:3dcartan2} are satisfied provided 
that the following conditions are met
\begin{gather}
\label{eq:hypersol_der_W}
 W'(r)=-\frac{g}{\sqrt{2}}\PP_\Lambda{}^3\I^\Lambda=
 -\frac{g}{\sqrt{2}}\left(c\,\I^0+\frac{e^{2\phi}}{2}\I^\Lambda k_\Lambda\right)\,,\\
\nonumber\\
\label{eq:hypersol_A0}
 \tilde A^0=-\frac{f'(\vartheta)}{gc}d\varphi\,.
\end{gather}
From \eqref{eq:hypersol_A0} and the constraint \eqref{eq:hypersol_gauge_constr} we also have
\begin{equation}
 \label{eq:hypersol_A1}
 \tilde A^1=\frac{k_0}{k_1}\frac{f'(\vartheta)}{gc}d\varphi\,.
\end{equation}
Finally, \eqref{eq:ftilde} leads to the two equations
\begin{equation}
\label{eq:hypersol_Ftilde}
 \left[\left(\I^\Lambda k_\Lambda \right)'-\frac{g}{\sqrt{2}}\left(\I^\Lambda\right)^2 k_\Lambda\PP_\Lambda{}^3\right]e^{2 W(r)}
 =(-1)^\Lambda \frac{\sqrt{2} k_0}{gc}\frac{f''(\vartheta)}{f(\vartheta)}
\quad\mbox{(no sum over $\Lambda$)}\,,
\end{equation}
while \eqref{eq:Iintegrab} is automatically satisfied since we obtained $\tilde F^\Lambda$ as the exterior derivative of the 
effective connection $\tilde A^\Lambda$.

Equation \eqref{eq:hypersol_derphi} allows us to use the chain rule to trade the coordinate $r$ for $\phi$ in 
\eqref{eq:hypersol_Ftilde}, which after summing over $\Lambda$ becomes
\begin{equation}
\frac12 \partial_\phi\left[\left( \I^\Lambda k_\Lambda \right)^2  \right] -\left( \I^\Lambda k_\Lambda \right)^2
+2\,\I^0 k_0\left( \I^1 k_1-\I^0 c\, e^{-2\phi} \right)=0\,.
\end{equation}
If we impose the condition
\begin{equation}
\label{eq:hypersol_impose}
 \I^1 k_1=\I^0c\,e^{-2\phi}\,,
\end{equation}
 this equation is solved by
\begin{equation}
 \label{eq:hypersol_I0_I1}
 \I^0=\frac{\alpha e^\phi}{k_0+c\,e^{-2\phi}}\,,\qquad\I^1=\frac{c}{k_1}\frac{\alpha e^{-\phi}}{k_0+c\,e^{-2\phi}}\,,
\end{equation}
where $\alpha$ is an integration constant. Substituting these expressions back in \eqref{eq:hypersol_Ftilde} for $\Lambda=0$ or 
$\Lambda=1$, we obtain an expression for the function $W(r)$,
\begin{equation}
\label{eq:hypersol_W}
 e^{2 W(r)}=\left[\frac{2}{\alpha g c}\left(k_0+c\,e^{-2\phi} \right)e^{-\phi}  \right]^2
\frac{f''(\vartheta)}{f(\vartheta)}\,.
\end{equation}
The expression \eqref{eq:hypersol_W} is also a solution of equation \eqref{eq:hypersol_der_W}, which is non-trivial, proving
the constraint \eqref{eq:hypersol_impose} to be consistent with all the equations. From
\eqref{eq:hypersol_W} we also conclude that $f''(\vartheta)/f(\vartheta)$ should be a positive constant,
therefore $f(\vartheta)$ in general takes the form
\begin{equation}
\label{eq:hypersol_f}
 f(\vartheta)=\gamma\sinh\left( \delta\vartheta+\rho \right)\,,
\end{equation}
where $\gamma$, $\delta$ and $\rho$ are constants.
We can now go back to the coordinate $r$ by solving equation \eqref{eq:hypersol_derphi} to obtain the dependence of $\phi$ on $r$, 
obtaining
\begin{equation}
 \phi=-\frac13 \log\left( -\frac{3 \alpha g}{2\sqrt{2}}r+\beta \right)\,,
\end{equation}
where $\beta$ is yet another integration constant. Note that all the integration constants can be
reabsorbed by the coordinate change
\begin{equation}
\label{eq:hypersol_coord_change}
 (\,t\,,\,r\,,\,\vartheta\,,\varphi\,)\quad\longrightarrow\quad
   \left(\,\frac{4\sqrt{2}\,\alpha}{g k_1 c} \,t\,,\,-\frac{2\sqrt{2}}{3\alpha g}\left(r^3-\beta  \right)\,,\,
            \frac{\vartheta-\rho}{\delta}\,,\,\frac{\varphi}{\delta\gamma}\,\right)\,,
\end{equation}
that allows to write the complete solution as
\begin{gather}
\label{eq:hypersol_fin_metric}
 ds^2=\frac{16\,r^2}{g^2 k_1 c}\!\left[\left(1+\frac{k_0}{c}\frac{1}{r^2} \right)^{\!\!2}\! r^2 dt^2
 -\!\left(1+\frac{k_0}{c}\frac{1}{r^2} \right)^{\!\!-2}\!\! \frac{dr^2}{r^2}-
 \frac{1}{2}\left( d\vartheta^2\!+\sinh^2\vartheta \,d\varphi^2 \right)\right]\,,\\
 \nonumber\\
\label{eq:hypersol_fin_gauge}
 A^0=-\frac{\cosh\vartheta}{gc}d\varphi\,,\qquad A^1=\frac{k_0}{k_1}\frac{\cosh\vartheta}{gc}d\varphi\,,\\
 \nonumber\\
 \label{eq:hypersol_fin_scalars}
 \phi=-\log r\,,\qquad  Z=\frac{c}{k_1}\, r^2\,.
\end{gather}
We start the analysis of the solution by noting that it has no free parameters, since all the constants appearing in 
\eqref{eq:hypersol_fin_metric}--\eqref{eq:hypersol_fin_scalars} are completely determined by the choice of gauging.
Observe also that in order to maintain the correct signature and to have $Z>0$, which is required to have a real K\"ahler potential, we 
have to impose $k_1 c>0$. 

The metric \eqref{eq:hypersol_fin_metric} is singular in $r=0$ and, if $k_0 c<0$, also in $r=\sqrt{-k_0/c}$. The singularity
in $r=r_S\equiv 0$ is a true curvature singularity, while the one in $r=r_H\equiv\sqrt{-k_0/c}$ is not and corresponds instead to a 
Killing horizon, always covering the curvature singularity.

With the metric written in the form \eqref{eq:hypersol_fin_metric}, it is immediate to see that in the asymptotic limit 
$r\rightarrow +\infty$ it reduces to
\begin{equation}
\label{eq:hypersol_asympt_metric}
 ds^2=\frac{16 r^2}{g^2 k_1 c}\left[r^2 dt^2-\frac{dr^2}{r^2}
 -\frac{1}{2}\left( d\vartheta^2 +\sinh^2\vartheta \,d\varphi^2 \right)\right]\,,
\end{equation}
which is manifestly conformally equivalent to AdS$_2\times$H$^2$. Note that
\eqref{eq:hypersol_asympt_metric} is very similar to hyperscaling violating geometries, which in $d$
dimensions have the form
\begin{equation}
ds^2 = r^{-\frac{2\theta}{d-2}}\left(r^{2z}dt^2 - \frac{dr^2}{r^2} - r^2(dx^i)^2\right)\,, \label{hyp-viol}
\end{equation}
where $i=1,\ldots,d-2$. Here, $z$ is the dynamical critical exponent and $\theta$ the so-called
hyperscaling violation exponent. Under the scaling $r\to r/\lambda$, $x^i\to\lambda x^i$,
$t\to\lambda^z t$, \eqref{hyp-viol} is not invariant, but transforms covariantly,
$ds\to\lambda^{\theta/(d-2)}ds$. Geometries of the form \eqref{hyp-viol} have been instrumental
in recent applications of AdS/CFT to condensed matter physics, cf.~e.g.~\cite{Huijse:2011ef}.
\eqref{eq:hypersol_asympt_metric} exhibits actually a scaling behaviour similar to that of
\eqref{hyp-viol}. To see this, introduce new coordinates $x$, $y$ on $\text{H}^2$ according to
\begin{equation}
x + iy = \frac{e^{i\varphi}\tanh\frac{\vartheta}2 + 1}{e^{i\varphi}\tanh\frac{\vartheta}2 - 1}\,,
\end{equation}
which casts \eqref{eq:hypersol_asympt_metric} into the form
\begin{equation}
ds^2 = \frac{16 r^2}{g^2 k_1 c}\left[r^2 dt^2-\frac{dr^2}{r^2} - \frac{dx^2 + dy^2}{2x^2}\right]\,.
\label{hyp-viol-prime}
\end{equation}
Under the scaling
\begin{equation}
r \to \frac r\lambda\,, \qquad t \to \lambda t\,, \qquad x \to \lambda x\,, \qquad y \to \lambda y\,,
\end{equation}
\eqref{hyp-viol-prime} transforms as $ds\to ds/\lambda$.

In the near-horizon limit, $r\rightarrow r_H$, after the coordinate change $t\rightarrow t/4$, the metric takes the form
\begin{equation}
 ds^2=-\frac{4}{g^2c^2}\frac{k_0}{k_1}\left[r^2 dt^2-\frac{dr^2}{r^2}-2\left( d\vartheta^2 +
\sinh^2\vartheta\, d\varphi^2 \right)\right]\,,
\end{equation}
which is AdS$_2\times$H$^2$, while the scalar fields take the values
\begin{equation}
 \phi=-\frac12 \log\left( -\frac{k_0}{c} \right)\,,\qquad\qquad Z=-\frac{k_0}{k_1}\,.
\end{equation}

The magnetic charges are given by
\begin{equation}
 P^\Lambda=\frac{1}{4\pi}\int F^\Lambda=p^\Lambda \mathbf{V}\,,\qquad \mathbf{V}=\int\sinh\vartheta
 \,d\vartheta\wedge d\varphi\,,
\end{equation}
yielding for the magnetic charge densities
\begin{equation}
 p^0=-\frac{1}{4\pi g c}\,\qquad p^1=\frac{k_0}{k_1}\frac{1}{4\pi g c}\,.
\end{equation}
The Bekenstein-Hawking entropy density can then be written as
\begin{equation}
 s=\frac{S}{\mathbf{V}}=-\frac{k_0}{k_1}\frac{2}{g^2c^2}=32\pi^2 p^0 p^1\,.
\end{equation}

\section{Attractor mechanism}
\label{sec:attractor}

In \cite{Bellucci:2008cb} the authors presented a generalization of the well-known black hole attractor mechanism
\cite{Ferrara:1995ih,Strominger:1996kf,Ferrara:1996dd,Ferrara:1996um,Ferrara:1997tw} to extremal static black holes in $N=2$, $d=4$ gauged 
supergravity coupled to abelian vector multiplets. In this section we closely follow their argument, generalizing it by taking into 
account the presence of gauged hypermultiplets. As in \cite{Bellucci:2008cb}, we make no assumption on the form of the scalar potential, 
of the vectors' kinetic matrix \ma{N} or on the scalar manifolds, so that our results are valid not only for $N=2$ supergravity, but for 
any theory described by an action of the form \eqref{eq:action}.

The equations of motion obtained from the variation of \eqref{eq:action} are
\begin{gather}
 \label{eq:att_eineq}
 R_{\mu\nu}+T_{\mu\nu}+2\ma{G}_{i\bar\jmath}\,\partial_{(\mu}Z^i\partial_{\nu)}\bar Z^{\bar\jmath}
 +2\masf{H}_{uv}\mathfrak{D}_{\mu}q^u\mathfrak{D}_{\nu}q^v-\frac12 g_{\mu\nu}V=0\,, \\
 \nonumber\\
 \label{eq:att_maxeq}
 \nabla_\nu\left( \star F_\Lambda{}^{\nu\mu} \right)+\frac g2\kk_{\Lambda u}\mathfrak{D}^\mu q^u=0\,,\\
 \nonumber\\
 \label{eq:att_cscaleq}
 \mathfrak{D}^2 Z^i+\partial^i F_\Lambda{}^{\mu\nu}\star F^\Lambda{}_{\mu\nu}+
 \frac12\partial^i V=0\,, \\
 \nonumber\\
 \label{eq:att_hscaleq}
 \mathfrak{D}^2 q^u+\frac14 \partial^u V=0\,,
\end{gather}
where
\begin{equation}
 T_{\mu\nu}\equiv I_{\Lambda\Sigma}\left( 4 F^{\Lambda\phantom{\mu}\rho}_{\phantom{\Lambda}\mu} F^\Sigma_{\phantom{\Sigma}\nu\rho}
                  -g_{\mu\nu}F^{\Lambda}_{\phantom{\Lambda}\rho\sigma} F^{\Sigma\rho\sigma}\right)\,,
\end{equation}
the dual field strengths are given by
\begin{equation}
 \label{eq:att_dualF}
 F_{\Lambda\mu\nu}\equiv-\frac{1}{4\sqrt{|g|}}\frac{\delta S}{\delta \star F^{\Lambda\mu\nu}}=
 R_{\Lambda\Sigma} F^{\Sigma}_{\phantom{\Sigma}\mu\nu}+I_{\Lambda\Sigma}\star F^{\Sigma}_{\phantom{\Sigma}\mu\nu}\,,
\end{equation}
and the second covariant derivatives on the scalars act as
\begin{gather}
 \mathfrak{D}^2 Z^i=\nabla_\mu \partial^\mu Z^i+\Gamma^i_{jk}\partial_\mu Z^j\partial^\mu Z^k\,, \\
 \nonumber\\
 \mathfrak{D}^2 q^u=\nabla_\mu \mathfrak{D}^\mu q^u+\Gamma^u_{vw}\mathfrak{D}_\mu q^v \mathfrak{D}^\mu q^w
                     +g A^\Lambda{}_\mu \partial_v \kk_\Lambda{}^u\mathfrak{D}^\mu q^v\,.
\end{gather}
The metric for the most general static extremal black hole background with flat, spherical or hyperbolic horizon can be
written in the form
\begin{equation}
 ds^2=e^{2 U(r)}dt^2-e^{-2 U(r)}\left[ dr^2 +e^{2 W(r)}\left( d\vartheta^2+f_\kappa
 (\vartheta)^2 d\varphi^2 \right)\right]\,,
\end{equation}
with 
\begin{equation}
 f_\kappa (\vartheta)=\left\{\begin{array}{r@{\quad}l}
                    \sin\vartheta\,,  & \kappa=1\,, \\
                    \vartheta\,,      & \kappa=0\,, \\
                    \sinh\vartheta\,, & \kappa=-1\,.\\
                   \end{array}\right.
\end{equation}
We require that all the fields are invariant under the symmetries of the metric, namely the time translation isometry generated by 
$\partial_t$ and the spatial isometries generated by the Killing vectors
\begin{equation}
 \partial_\varphi\,, \qquad \cos\varphi\,\partial_\vartheta-\frac{f_\kappa'}{f_\kappa}\sin\varphi\,
 \partial_\varphi\,, \qquad 
 \sin\varphi\,\partial_\vartheta+\frac{f_\kappa'}{f_\kappa}\cos\varphi\,\partial_\varphi\,.
\end{equation}
The scalar fields can then only depend on the radial coordinate $r$, and the request of invariance of the field strength 2-forms
$F^\Lambda$ leads to
\begin{equation}
 F^\Lambda=\frac12 F^\Lambda{}_{\mu\nu}(x)dx^\mu dx^\nu= 
 F^\Lambda{}_{tr}(r) dt\wedge dr+ F^\Lambda{}_{\vartheta\varphi}(r,\vartheta) d\vartheta\wedge d\varphi\,,
\end{equation}
with
\begin{equation}
\label{eq:att_magnF}
 F^\Lambda{}_{\vartheta\varphi}(r,\vartheta)=4\pi p^\Lambda (r) f_\kappa (\vartheta)\,,
\end{equation}
where $p^\Lambda (r)$ is a generic function of $r$. The Bianchi identities
\begin{equation}
 \nabla_\nu\left( \star F^{\Lambda\nu\mu} \right)=0\qquad \Longleftrightarrow \qquad \partial_{[\mu}F^\Lambda{}_{\nu\rho]}=0
\end{equation}
imply that $p^\Lambda$ must be constant. With field strengths of this form, it is always possible to choose a gauge in which the gauge potential 1-forms can be written as
\begin{equation}
 A^\Lambda=A^\Lambda{}_t (r) dt+A^\Lambda{}_\varphi (\vartheta) d\varphi\,.
\end{equation}
The $r$-component of the Maxwell equations \eqref{eq:att_maxeq} reduces then to the condition
\begin{equation}
 \kk_{\Lambda u}(q) \partial_r q^u=0\,,
\end{equation}
while the $\vartheta$-component is automatically satisfied and the $\varphi$-component gives 
\begin{equation}
 A^\Sigma{}_\varphi\kk_\Sigma{}^u\kk_\Lambda{}_u=0
\end{equation}
for every value of $\Lambda$, or equivalently
\begin{equation}
 \kk_\Lambda{}^u (q)\, p^\Lambda=0\,.
\end{equation}
Finally if we define a function $e_\Lambda (r)$ such that 
\begin{equation}
\label{eq:att_elF}
 F^\Lambda{}_{tr}(r)=4\pi I^{\Lambda\Sigma}\left( e_\Sigma(r)-R_{\Sigma \Gamma} p^\Gamma \right)e^{2(U-W)}\,,
\end{equation}
we have $F_{\Lambda\vartheta\varphi}=4\pi e_\Lambda (r) f_\kappa(\vartheta)$ and the $t$-component
of the Maxwell equations becomes
\begin{equation}
\label{eq:att_maxw_t}
 4\pi e^{2(U-W)}\partial_r e_\Lambda=\frac{g^2}{2}e^{-2 U} A^\Sigma{}_t\kk_\Sigma{}^u\kk_\Lambda{}_u\,.
\end{equation}
The quantities $p^\Lambda$ and $e_\Lambda(r)$ are the magnetic and electric charge densities inside the 2-surfaces $S_r$ of constant $r$ and $t$,
\begin{equation}
 p^\Lambda=\frac{1}{4\pi \mathbf{V}}\int_{S_r} F^\Lambda\,,\qquad e_\Lambda(r)=\frac{1}{4\pi \mathbf{V}}\int_{S_r} F_\Lambda\,,\qquad 
 \mathbf{V}=\int_{S_r} f_\kappa (\vartheta)d\vartheta\wedge d\varphi\,.
\end{equation}
The non-vanishing components of $T_{\mu\nu}$ are given by
\begin{equation}
 T_t^t=T_r^r=-T_\theta^\theta=-T_\varphi^\varphi=(8\pi)^2 e^{4(U-W)} \tilde V_{\text{BH}}\,,
\end{equation}
where $\tilde V_{\text{BH}}$ is the so-called black hole potential,
\begin{equation}
 \tilde V_{\text{BH}}=-\frac12 \begin{pmatrix}
                  p^\Lambda\,, & e_\Lambda(r)
                 \end{pmatrix}
                 \begin{pmatrix}
                  I_{\Lambda\Sigma}+R_{\Lambda\Gamma}I^{\Gamma\Omega}R_{\Omega\Sigma}\: & \:-R_{\Lambda\Gamma}I^{\Gamma\Sigma} \\
                  -I^{\Lambda\Gamma}R_{\Gamma\Sigma}\:                                  & \:I^{\Lambda\Sigma}
                 \end{pmatrix}
                 \begin{pmatrix}
                  p^\Sigma \\
                  e_\Sigma (r)
                 \end{pmatrix}\,,
\end{equation}
which however, unlike the usual definition, has an explicit dependence on $r$ through the varying electric charges $e_\Lambda$.
It is also straightforward, using the expressions \eqref{eq:att_magnF}, \eqref{eq:att_elF} and the definition \eqref{eq:att_dualF}, to 
verify that
\begin{equation}
 \partial^i F_\Lambda{}^{\mu\nu}\star F^\Lambda{}_{\mu\nu}=(8\pi)^2 e^{4(U-W)}\partial^i
 \tilde V_{\text{BH}}\,,
\end{equation}
where on the left-hand side only the dual field strengths $F_\Lambda$ are taken to depend on the complex scalars $Z^i$ and only through
the matrices $R_{\Lambda\Sigma}$ and $I_{\Lambda\Sigma}$ appearing in \eqref{eq:att_dualF}, while on the right-hand side the charges 
$e_\Lambda(r)$ are considered to be independent of the $Z^i$.
Equations (\ref{eq:att_eineq}), (\ref{eq:att_cscaleq}) and (\ref{eq:att_hscaleq}) then reduce to
\begin{gather}
\label{eq:att_symm_eom_eintt}
 e^{2 U}\left( 2 U' W'+U'' \right)-(8\pi)^2 e^{4(U-W)}\tilde V_{\text{BH}}-2 g^2 e^{-2U}A^\Lambda{}_t\kk_\Lambda{}_uA^\Sigma{}_t\kk_\Sigma{}^u
  +\frac{V}{2}=0\,, \\
 \nonumber\\
\label{eq:att_symm_eom_einrr}
 e^{2 U}\left(U'^2+ W'^2+ W'' \right)-(8\pi)^2 e^{4(U-W)}\tilde V_{\text{BH}}+ e^{2U}\ma{G}_{i\bar\jmath}Z^{i\prime}{\bar Z}^{\bar\jmath\prime}
  + e^{2U}\masf{H}_{uv}q^{u\prime}q^{v\prime}\nonumber\\
  \qquad\qquad\qquad\qquad\qquad\qquad\qquad\qquad\qquad-g^2 e^{-2U}A^\Lambda{}_t\kk_\Lambda{}_uA^\Sigma{}_t\kk_\Sigma{}^u+\frac{V}{2}=0\,, \\
 \nonumber\\
\label{eq:att_symm_eom_einthth}
 e^{2 U}\left(-\kappa e^{-2 W} +2 W'^2+W'' \right)-2 g^2 e^{-2U}A^\Lambda{}_t\kk_\Lambda{}_uA^\Sigma{}_t\kk_\Sigma{}^u+V=0\,, \\
 \nonumber\\
\label{eq:att_symm_eom_Z}
 e^{2 U}\left(Z^{i\prime\prime}+2 W' Z^{i\prime}+ \ma{G}^{i\bar\jmath}\partial_l\ma{G}_{k\bar\jmath}Z^{l\prime}Z^{k\prime}\right)
  -(8\pi)^2 e^{4(U-W)}\partial^i \tilde V_{\text{BH}}-\frac12\partial^i V=0\,, \\
 \nonumber\\
\label{eq:att_symm_eom_q}
 e^{2 U}\left(q^{u\prime\prime}+2 W' q^{u\prime}+ \Gamma^u_{vz}q^{v\prime}q^{z\prime}\right)-
 g^2 e^{-2U}A^\Lambda{}_t\kk_\Lambda{}^v A^\Sigma{}_t \nabla_v\kk_\Sigma{}^u  -\frac14 \partial^u V=0\,,
\end{gather}
where a prime denotes a derivative with respect to $r$.

In the near horizon limit ($r\rightarrow 0$) one has
\begin{equation}
\label{eq:att_metric_limits}
 U\sim \log \frac{r}{r_{\text{AdS}}}\,,\qquad W\sim\log\left( \frac{r_H}{r_{\text{AdS}}} r \right)\,,
\end{equation}
where $r_{\text{AdS}}$ is the $\text{AdS}_2$ curvature radius.
We require all the fields, their derivatives, the scalar potential and the couplings to be regular on the horizon. Then we can choose a 
gauge such that
\begin{equation}
\label{eq:att_At_gauge}
 \left.A^\Lambda{}_t\right|_{r=0}=0\qquad\Longrightarrow\qquad A^\Lambda{}_t\stackrel{r\to 0}{\sim}\left.F^\Lambda{}_{rt}\right|_{r=0} r\,.
\end{equation}
It is also reasonable to assume that the derivative of the electric charges $\partial_r e_\Lambda$
remains finite on the horizon. In this case, 
eq. \eqref{eq:att_maxw_t} implies that in the near-horizon limit the quantity $A^\Sigma{}_t\kk_\Sigma{}_u\kk_\Lambda{}^u$ is at least of
order $r^2$. If we expand in powers of $r$, in the gauge \eqref{eq:att_At_gauge} the order zero term automatically vanishes, while for the 
order one term we have
\begin{equation}
\label{eq:att_Fk}
 0=\left.\partial_r \left(A^\Sigma{}_t\kk_\Sigma{}_u\kk_\Lambda{}^u\right)\right|_{r=0}=
 \left.-F^\Sigma{}_{tr}\kk_\Sigma{}_u\kk_\Lambda{}^u\right|_{r=0}
 \qquad\Longrightarrow\qquad \left.F^\Lambda{}_{tr}\kk_\Lambda{}^u\right|_{r=0}=0\,.
\end{equation}
Using \eqref{eq:att_At_gauge} and \eqref{eq:att_Fk} one can see that the terms with $A^\Lambda{}_t$ in the equations of motion, 
$e^{-2U}A^\Lambda{}_t\kk_\Lambda{}_uA^\Sigma{}_t\kk_\Sigma{}^u$  and 
$e^{-2U}A^\Lambda{}_t\kk_\Lambda{}^v A^\Sigma{}_t \nabla_v\kk_\Sigma{}^u$, go to zero in the
near-horizon limit.
In this limit the equations of motion \eqref{eq:att_symm_eom_eintt}--\eqref{eq:att_symm_eom_q} thus reduce to
\begin{gather}
 \frac{1}{r_{\text{AdS}}^2}=(8\pi)^2 \frac{V_{\text{BH}}}{r_H^4}-\frac{V}{2}\,, \\
 \nonumber\\
 \frac{\kappa}{r_H^2}=\frac{1}{r_{\text{AdS}}^2}+V\,, \\
 \nonumber\\
 \label{eq:att_dZ_Veff}
 \partial_i\left[ (8\pi)^2 \frac{V_{\text{BH}}}{r_H^4}+\frac{V}{2} \right]=0\,, \\
 \nonumber\\
 \label{eq:att_dq_Veff}
 \partial_u V=0\,,
\end{gather}
where $V_{\text{BH}}\equiv\tilde V_{\text{BH}}|_{e_\Lambda(r)\to e_\Lambda(0)}$. Solving the first two
equations for $r_H^2$ and $r_{\text{AdS}}^2$ one gets
\begin{gather}
 r_H^2=\left.\frac{\kappa\pm\sqrt{\kappa^2-2(8\pi)^2 V_{\text{BH}} V }}{V}\right|_{r=0}\,, \\
 \nonumber\\
 r_{\text{AdS}}^2=\left.\mp \frac{r_H^2}{\sqrt{\kappa^2-2(8\pi)^2 V_{\text{BH}} V }}\right|_{r=0}\,,
\end{gather}
and since of course $r_{\text{AdS}}^2>0$ we have to choose the lower sign. We also have to require
$r_H^2>0$, which means that flat or hyperbolic
geometries, $\kappa=0,-1$, are only possible if the scalar potential takes negative values on the horizon, $V|_{r=0}<0$. Spherical 
geometry ($\kappa=1$), on the other hand, is compatible with both positive or negative values of $V$ on the horizon, but for $V|_{r=0}>0$ 
there is the restriction $V_{\text{BH}} V|_{r=0}<\frac{1}{2(8\pi)^2}$, since $V_{\text{BH}}$ is always positive.

We can introduce an effective potential as a function of the scalars,
\begin{equation}
 V_{\text{eff}}(Z,\bar Z, q)\equiv \frac{\kappa-\sqrt{\kappa^2-2(8\pi)^2 V_{\text{BH}} V }}{V}\,,
\end{equation}
defined for $V_{\text{BH}} V<\frac{1}{2(8\pi)^2}$, and write
\begin{gather}
 r_H^2=\left.V_{\text{eff}}\right|_{Z_H,q_H}\,, \\
 \nonumber\\
 r_{\text{AdS}}^2=\left.\frac{V_{\text{eff}}}{\sqrt{\kappa^2-2(8\pi)^2 V_{\text{BH}} V}}\right|_{Z_H,q_H}\,,
\end{gather}
with $Z^i_H\equiv\lim_{r\to 0} Z^i$, $q^u_H\equiv\lim_{r\to 0} q^u$.
Because of equations \eqref{eq:att_dZ_Veff}--\eqref{eq:att_dq_Veff}, $V_{\text{eff}}$ is extremized on the horizon by all the scalar fields of the
theory,
\begin{equation}
\label{eq:att_Veff_extr}
 \left.\partial_i V_{\text{eff}}\right|_{Z_H,q_H}=0\,,\qquad \left.\partial_u V_{\text{eff}}\right|_{Z_H,q_H}=0\,.
\end{equation}

The values $Z^i_H, q^u_H$ of the scalars on the horizon are then determined by the extremization conditions \eqref{eq:att_Veff_extr},
and the Bekenstein--Hawking entropy density is given by the critical value of $V_{\text{eff}}$,
\begin{equation}
 s=\frac{S}{\mathbf{V}}=\frac{A}{4 \mathbf{V}}=\frac{r_H^2}{4}=\frac{V_{\text{eff}}(Z_H,\bar Z_H, q_H)}{4}\,.
\end{equation}
For a given theory this critical value, and thus also the entropy, depend only on the charges (on the horizon) $p^\Lambda$ and 
$e_\Lambda (0)$, so that the attractor mechanism still works. On the other hand $Z^i_H$ and $q^u_H$ may not be uniquely determined, since
in general $V_{\text{eff}}$ may have flat directions.

The limit for $V\to 0$ of $V_{\text{eff}}$ only exists for $\kappa=1$, in which case
$V_{\text{eff}}\to (8\pi)^2 V_{\text{BH}}$ and one recovers the attractor
mechanism for ungauged supergravity. The fact that this limit does not exist for $\kappa=0,-1$ is not surprising since flat or hyperbolic
horizon geometries are incompatible with vanishing cosmological constant.

For the black hole we presented in section \ref{sec:bhsol}, the fact that the entropy only depends on the charges is not really
surprising, since the solution has no free parameters at all. It is however straightforward to verify that the near-horizon geometry
does indeed extremize the effective potential $V_{\text{eff}}$. In particular one has on the horizon
\begin{equation}
 \partial_{q} V=\partial_Z V=\partial_Z V_{\text{BH}}=0\,.
\end{equation}

\section{Final remarks}
\label{sec:final}

In this paper, we considered $N=2$ supergravity in four dimensions, coupled to vector- and
hypermultiplets, where abelian isometries of the quaternionic K\"ahler manifold are gauged.
In the first part, we analytically constructed a magnetically charged supersymmetric black hole solution
of this theory for the case of just one vector multiplet with prepotential ${\cal F}=-i\chi^0\chi^1$, and
the universal hypermultiplet. This black hole has a running dilaton, and interpolates between 
$\text{AdS}_2\times\text{H}^2$ at the horizon and a hyperscaling-violating type geometry at
infinity, which is conformal to $\text{AdS}_2\times\text{H}^2$. To the best of our knowledge, this
represents the first example of an analytic genuine BPS black hole in gauged supergravity with
nontrivial hyperscalars; previously known solutions of this type were only constructed
numerically \cite{Halmagyi:2013sla}.

Diverging scalars fields of the form \eqref{eq:hypersol_fin_scalars}
are common in two and three dimensions, but are sometimes regarded as a sign of pathology in four
or higher dimensions. However, similar to the linear dilaton black holes of \cite{Chan:1995fr}, our
solutions have finite entropy, magnetic charges and curvature at large $r$, in spite of the diverging
scalars, and should thus be regarded as physically meaningful\footnote{$R$, $R_{\mu\nu}R^{\mu\nu}$
and $R_{\mu\nu\rho\sigma}R^{\mu\nu\rho\sigma}$ decay for large $r$ like $r^{-2}$, $r^{-4}$
and $r^{-4}$ respectively.}. In any case, it may be interesting to consider more general models
and gaugings,
and to look for asymptotically AdS black holes with running hyperscalars, that might be more relevant
for gauge/gravity duality applications. Unfortunately the equations of \cite{Meessen:2012sr} become
immediately quite involved once the complexity of the model increases, but perhaps our solution
may serve as a starting point that helps solving analytically the equations of \cite{Meessen:2012sr}
in a more complicated setting. We hope to come back to this point in a future publication.

In the second part of the paper, we extended the work of \cite{Bellucci:2008cb} on black hole attractors
in gauged supergravity to the case where also hypermultiplets are present. The attractors were shown
to be governed by an effective potential $V_{\text{eff}}$, which is extremized on the horizon by all the
scalar fields of the theory. Moreover, the entropy is given by the critical value of $V_{\text{eff}}$, and in
the limit of vanishing scalar potential, $V_{\text{eff}}$ reduces (up to a prefactor) to the usual black hole
potential. The resulting attractor equations \eqref{eq:att_Veff_extr} do not make use of supersymmetry;
they are valid for any static extremal black hole. It would be interesting to analyze them for some
specific models, for instance the ones worked out in \cite{Cassani:2012pj} and considered also
in \cite{Halmagyi:2013sla} that arise from M-theory compactifications.

\end{document}